\documentclass[preprint,showpacs]{revtex4}
\usepackage{dcolumn}
\usepackage{amssymb}
\usepackage{bm}
\usepackage{array}
\usepackage{graphicx}
\usepackage{caption2}
\usepackage{amsmath}
\usepackage{hyperref}
\usepackage{subfigure}
\usepackage{amsfonts}

\usepackage{ulem}
\usepackage{xcolor}

\begin{document}

\title{Control excitation and coherent transfer in a dimer}
\author{Hong-rong Li$^{1}$, Pei Zhang$^{1}$, Yingjun Liu$^{2}$, Fu-li Li$^{1}$, and Shi-yao Zhu$^{3}$}
\affiliation{$^{1}$Department of Applied Physics, Xian Jiaotong University, Xian 710049, China}
\affiliation{$^{2}$The High School Affiliated to Xi'an Jiaotong University, Xian 710054, China}
\affiliation{$^{3}$Beijing Computational Science Research Center, Beijing 100084, China}

\begin{abstract}
In this article, the processes of energy absorption and coherent transfer in a dimer is studied. The dimer includes two two-level pigments --- donor and acceptor, where donor is assumed being excited by a control pulse in the time domain. We investigate the dynamics of probability that the acceptor is in the excited state and the total efficiency of energy absorption and transfer under different temporal shape of control pulse. Quantum concurrence of the dimer is also discussed.\\

\end{abstract}

\maketitle
\section{Introduction}

The primary processes in photosynthesis have been paid much interest \cite{[1],[2],[3],[4],[5],[6],[7]} from a broader physical community in recent years, thanks to experimental observation via electronic spectroscopy technology \cite{[8]} demonstrated that quantum coherence is involved in the excitation energy transfer of the light-harvesting complexes \cite{[3]} and Fenna-Matthews-Olson complex \cite{[9]}. In most of the photosynthetic processes, photochemical excitation of an antenna molecule by absorbing a pulse of light takes place firstly, and the absorbed excitation energy is then transferred among molecules of the photosynthetic systems until reaction centers where the energy is converted into chemical energy \cite{[1],[2],[3]}. 

 According to the F\"orster theory \cite{[10]}, when the electronic coupling between pigments is small in comparison to the electron-environment coupling, the energy transfer between different pigments takes place through incoherent hopping, where the electronic coupling can be treated perturbatively. On the contrary, when the electronic coupling between pigments is similar or larger than reorganization energy of the pigments, electronic excitations then move coherently through different pigments rather than by incoherent hopping motion \cite{[11]}. In the later case, the electron-environment coupling can be treated perturbatively to obtain a quantum master equation \cite{[12]}. Recent experimental~\cite{[13],[14],[15],[16],[17]} and theoretical works \cite{[18],[19],[20],[21],[22],[23],[24],[25],[26],[27],[28],[29],[30],[31],[32],[33],[34],[35],[36],[37],[38],[39]} support the coherent transfer case and indicate that long-lasting electronic coherence can indeed influence the excitation-transfer dynamics in photosynthetic complexes. The process of energy transfer takes only a few hundred picoseconds and is performed with extraordinarily high efficiency \cite{[3]}. In the most of the above theoretical works, investigation has been focused only on the energy transfer processes with assumption that the initial pigments are in their excited states. However, this assumption is just possible only when the shape of light pulses is very sharp, $i.e.$ nearly a $\delta$ function pulse. In this situation, the process of light absorption occurs rapidly before excitation energy transferring starts. In fact, the shape of the pulses that antenna molecule absorbed may has temporal width, which means that the processes of energy transferring always take place at the same time with the processes of absorbing light pulse. Therefore, it is necessary to consider the effects of the shape of pulses on efficiency of energy transfer. This is the key motivation of our present work.

To investigate how the shape of light pulses affects the processes of energy absorption and transferring, in stead of considering a complicated network of pigments where the practical transfer processes take place, we will study a basic physical part to obtain the physical mechanism: a dimer system which consists of a donor pigment and an acceptor pigment modeled by two two-level systems. With assumption that the two pigments are both in their ground states initially, we will study the dynamics of dimer system after the donor pigment is excited by a light pulse and absorbs energy of light. Efficiency of energy absorption and transfer will be discussed by calculating the probabilities of the acceptor pigment in its upper state. Our study is suitable for quantum control settings under artificial laser light condition. The previous important studies in molecular dimers and excitation with coherent pulse can be found in \cite{39-1,39-2}.

The paper is organized as follows. In Sec. II, a theoretical model and simple analyses are presented. In Sec. III, numerical results are shown. Conclusions and final remarks are presented in Sec. IV.

\section{The theoretical model and the associated dynamics}
The free Hamiltonian of the two pigments is
\begin{eqnarray}
H_{1}=\frac{1}{2}\omega_1\sigma_z^{(1)}+\frac{1}{2}\omega_2\sigma_z^{(2)},
\end{eqnarray}
and the coupling Hamiltonian between the two pigments is given by
\begin{eqnarray}
H_2=J(\sigma_+^{(1)}\sigma_-^{(2)}+\sigma_-^{(1)}\sigma_+^{(2)}),
\end{eqnarray}
where $\omega_i$ represents energy separation of the $i$th pigment, $J$ is the coupling strength, and $\sigma_z^{(i)}=|e\rangle_i\langle e|-|g\rangle_i\langle g|$ is the Pauli operator for the $ith$ pigment, $\sigma_+^{(i)}=|e\rangle_i\langle g|$ and $\sigma_-^{(i)}=|g\rangle_i\langle e|$ are the arising and lowering operators for the $ith$ pigment, respectively. We assume that the donor pigment is excited by an external pulse, the associated Hamiltonian is 
\begin{eqnarray}
H_3=E(t)\sigma_+^{(1)}+E^*(t)\sigma_-^{(1)},
\end{eqnarray}
where $E(t)=E_{\Omega}(t)e^{i\Omega t}$ is a time dependent amplitude of the external pulse. For example, for the Gaussian type laser pulse, $E_{\Omega}(t)=\frac{E_0}{\sqrt{2\pi}\tau_p}e^{\frac{-t^2}{2\tau_p^2}}$, and $\tau_p$ is the full width half maximum of the pulse. $H_1$, $H_2$, and $H_3$ are the Hamiltonian of the two-level system, which can be simply denoted by $H_M=H_1+H_2+H_3$.
In real photosynthetic systems, the effect of noise from environment ($e.g.$ vibrational modes of protein molecules in environment) is unavoidable. Here, we naturally use a Bose bath to denote the environmental modes, and the coupling Hamiltonian of the system and the environmental modes is
\begin{eqnarray}
H_{MB}
&=&\sum_{j=1}^2 \sigma_z^{(j)}\sum_{k_j}g_{kj}(a^\dag_{k_j}+a_{k_j}),
\end{eqnarray}
where $a^\dag_{k_j}$ is the creation operator of the Bose bath with mode $k_j$, and $g_{k_j}$ is the coupling strength between the $j$th pigment and the mode $k_j$ of the bath. The free Hamiltonian of the bath is
\begin{eqnarray}
H_B=\sum_{k_j}\nu_{k_j}a_{k_j}^\dag a_{k_j},
\end{eqnarray}
where $\nu_{k_j}$ represents the frequency of the mode $k_j$.

To get the evolution of the system, we first write the Hamiltonian into its eigenspace. The eigen-equation of the systems is given by $H_M|\epsilon_j\rangle=\epsilon_j|\epsilon_j\rangle$, where $|\epsilon_j\rangle$ is the corresponding eigenvector for the $j$th eigenvalue $\epsilon_j$. With solving the corresponding secular equation $|H_M-\epsilon I|=0$, we obtain the four eigenvalues, $\epsilon_{1,2}=\mp\frac{1}{2}\sqrt{\varepsilon_1+2\varepsilon_0},~\epsilon_{3,4}=\pm\frac{1}{2}\sqrt{\varepsilon_1-2\varepsilon_0}$, where $\varepsilon_0=\sqrt{4|E|^2(J^2+\omega_2^2)+(J^2-\omega_1\omega_2)^2}$, and $\varepsilon_1=2J^2+4|E|^2+\omega_1^2+\omega_2^2$. We notice that $\epsilon_2\geq\epsilon_3\geq\epsilon_4\geq\epsilon_1$.  For each eigenvalue $\epsilon_i$, the corresponding eigenvector $|\epsilon_i\rangle$ is linear superposition of the four bare states $|\eta_1\rangle=|ee\rangle$, $|\eta_2\rangle=|eg\rangle$, $
|\eta_3\rangle=|ge\rangle$, and $|\eta_4\rangle=|gg\rangle$ of the two-level system. If we let $|\epsilon\rangle=[|\epsilon_1\rangle, |\epsilon_2\rangle, |\epsilon_3\rangle, |\epsilon_4\rangle]^T$, and $|\eta\rangle=[|\eta_1\rangle, |\eta_2\rangle, |\eta_3\rangle, |\eta_4\rangle]^T$, and use $U=\{u_{ij}\}$ to denote the transform matrix from $|\epsilon\rangle$ to $|\eta\rangle$, then we have a simple form between the original state vectors and the eigenstate vectors
\begin{eqnarray}
|\eta\rangle=U|\epsilon\rangle.
\end{eqnarray}
In the eigenspace, the diagonal form of the Hamiltonian is
\begin{eqnarray}
H_M=\sum_{i=1}^4 \epsilon_i |\epsilon_i\rangle\langle\epsilon_i|,
\end{eqnarray}
In the new basis, the Pauli operators $\sigma_z^{(m)}=|e\rangle_m\langle e|-|g\rangle_m\langle g|=\sum_{i,j=1}^4 s_{ij}^{(m)} |\epsilon_i\rangle\langle\epsilon_j|$, where $s_{ij}^{(1)}=u_{1i} u_{1j}^*+u_{2i} u_{2j}^*-u_{3i} u_{3j}^*-u_{4i} u_{4j}^*$, and $s_{ij}^{(2)}=u_{1i} u_{1j}^*-u_{2i} u_{2j}^*+u_{3i} u_{3j}^*-u_{4i} u_{4j}^*$. Using these notations, we rewrite the coupling Hamiltonian as
\begin{eqnarray}
H_{MB}=\sum_{l=1}^2\sum_{k_l}\sum_{i,j=1}^4 s_{ij}^{(l)} g_{k_l} |\epsilon_i\rangle\langle\epsilon_j| (a_{k_l}^{\dag}+a_{k_l}).
\end{eqnarray}

To give an elementary view of the dynamics of our model, we first consider a closed evolution based on the $H_M$. With assuming that the two pigments are both in their ground states initially, $|\varphi(0)\rangle=|gg\rangle=|\eta_4\rangle=\sum_{j=1}^4 u_{4j}(0)|\epsilon_{j}\rangle$, and after applying to Schr\"odinger equation, we have the formal solution of the system $|\varphi(t)\rangle=e^{-i\int H_M(t) dt}|\varphi(0)\rangle$. We consider two extreme cases: $~J^{-1}\gg\tau_p$, and $~J^{-1}\ll\tau_p$. 

In the first case of $~ J^{-1}\gg\tau_p\rightarrow 0$, the input pulse is a sharp wave packet, because that $a\rightarrow 0$, $\frac{1}{a \sqrt{\pi}}e^{-x^2/a^2}\rightarrow \delta(x)$. This means that for the Gaussian type pulse  $E_{\Omega}(t)\rightarrow E_0\delta(t)$ with $\sqrt{2}\tau_p\rightarrow 0$. We thus suppose that the whole dynamic process has two steps: 1. The donor pigment is exicted by the input pulse; 2. Excitation energy transfers from the donor to the acceptor. In the first step, we find $H_M \approx H_3$ as $E(t)\sim \delta(t)$, thus we have the dynamic states at time $t_1$ after the pulse takes action, $|\psi(t_1)\rangle = e^{-i\int_0^{t_1} H_3(t')dt'}|gg \rangle =e^{i\gamma_0}|eg \rangle$. After the action of $\delta(t)$ pulse, as $E(t)\sim 0$, we have $H_M \approx H_1+H_2$, which corresponding to the second step, $i.e.$ energy transferring from the donor pigment to the acceptor pigment starts when the donor pigment is in its excited state, which is also the common assumption in some of the published works about quantum dynamics of photosynthesis \cite{[28],[29],[30],[31],[32],[33]}.

The second extreme case associates to a near flat and continuous action pulse, with assuming simply that $E(t)\sim E_c$. We then find that $|\psi(t)\rangle \approx \sum_{j=1}^4 u_{4j}(0) e^{-i \epsilon_j t}|\epsilon_j \rangle$. Normally, the shape of a pulse absorbed by antenna molecule is not sharp --- the pulse has width in the space and time domain. Spatial and temporal coherence, and other effects coming from its shape should be considered, and the process of excitation transfer is surely affected by these effects because the spatial and temporal width of a control pulse is similar to the space and time scale in excitation transfer processes \cite{[13],[14],[15],[16],[17]}. On the one hand, as the wave packet of the photons captured by antenna pigments is always larger than or comparable with the scales of multi-chromophoric molecules~\cite{[6]}, the initial excitation takes place coherently among the antenna pigments. Thus efficiency of energy transferring from the antenna to the reaction center depends intimately on the quantum superposition properties of the initial states~\cite{[18]}, and these initial spatial coherence will enhance or trap transfer of the donor pigments at different conditions~\cite{[18],[19],[20],[21],[22]}. Similarly, on the other hand, under a single excitation assumption, a pulse with temporal width will induce excitation coherent at different time, $i.e.$ with different phase, the temporal shape of an input pulse can also affect processes of excitation transfer. For example, in a process of absorbing and transferring energy with a flat input pulse, the donor is firstly excited from $|gg\rangle$ to $|eg\rangle$ by the front part of the input pulse, and then coupling between two pigments induces excitation transfer from $|eg\rangle$ to $|ge\rangle$, the next part of the pulse will then coherently stimulates the pigments from $|gg\rangle$ to $|eg\rangle$ and from $|ge\rangle$ to $|ee\rangle$. It means the donor is then always in its excited states, which will lead to saturation of energy absorbing and transferring and thus increasing dissipation and reducing efficiency. Therefore, in a more realistic pulse absorbing process of photosynthesis, one needs to not only consider efficiency of energy transfer between different pigments, but also investigate the whole efficiency including pulse energy absorbing. To investigate the whole efficiency including pulse energy absorbing and energy transferring from the donor to the acceptor, we directly use the area under pulse figure in the time domain to denote the total power of the pulse, and define a parameter of total efficiency as following
\begin{eqnarray}
\eta_{total}=\frac{\omega_2 Tr[|e \rangle_2\langle e|]}{\int |E(t)|^2 dt}.
\end{eqnarray}
Note that the parameter of $\eta_{total}$ is not a true efficiency because it will be larger than 1 at some conditions. To compare the case without considering absorption processes, we will also draw figures of the probability that the acceptor is in its excited state, $i.e.$ $P=Tr[|e \rangle_2\langle e|]$.
Generally, the population in the donor pigment should be considered because the population correlates with the whole efficiency. However, in the present studies, the pigment-environment coupling is assumed smaller than the coupling between pigments, there are only resonant and near-resonant frequencies in the pulse should match the eigenstates which overlap with the donor pigment. So we will not consider the population in the donor pigment.

It is also important to study the dynamics of quantum entanglement of typical dimer systems. We choose the concurrence $C$ to quantify the entanglement \cite{[40]}, which is defined $C=\max \left\{0,\varepsilon _{0}\right\} $, and $\varepsilon _{0}=\varepsilon_{1}-\varepsilon _{2}-\varepsilon _{3}-\varepsilon _{4}$, where $\varepsilon_{i}$ are the square roots of the eigenvalues of $\rho \widetilde{\rho }$ in the decreasing order, and $\widetilde{\rho }=\left( \sigma _{y}\otimes
\sigma _{y}\right) \rho ^{\ast }\left( \sigma _{y}\otimes \sigma _{y}\right)$. 

\section{The master equation and numerical investigation}
In the eigenspace and under the second-order approximation of the coupling between the two-level systems and environment, the master equation is in the following form%
\begin{equation}
\dot{\rho}\left( t\right) =-i\left[ H_{M}\left( t\right) ,\rho \left(
t\right) \right] +L\rho \left( t\right) ,
\end{equation}%
where the Lindblad operator is $L\rho \left( t\right) =- \sum_{\mu =1}^{12}\xi_{\mu}\left( \{ \pi_{\mu}^{+}\pi_{\mu}, \rho \left( t\right)\ \}-2\pi_{\mu}\rho \left( t\right) \pi_{\mu}^{+} \right)$. The detailed parameters are given by%
\begin{eqnarray}
\xi _{m}=\left\{
\begin{array}{l}
\epsilon_{ij}\sum_{l=1}^2 \kappa_k s_{ij}^{(l)} s_{ji}^{(l)} \left[ N_l(\epsilon_{ij})+1\right];  ~m\leq 6 \\
\epsilon_{ij}\sum_{l=1}^2 \kappa_k s_{ij}^{(l)} s_{ji}^{(l)} N_l(\epsilon_{ij}) ;  ~7\leq m\leq 12%
\end{array}%
,\right.
\end{eqnarray}
where $\xi_m$ corresponding to all six level gaps greater than zero, $\epsilon_{21}$, $\epsilon_{23}$, $\epsilon_{24}$, $\epsilon_{31}$, $\epsilon_{34}$, and $\epsilon_{41}$, respectively.  And $\pi _{1}=\pi _{7}^{+}=\left\vert \epsilon _{3}\left( t\right)
\right\rangle \left\langle \epsilon _{2}\left( t\right) \right\vert$, $\pi
_{2}=\pi _{8}^{+}=\left\vert \epsilon _{4}\left( t\right) \right\rangle
\left\langle \epsilon _{2}\left( t\right) \right\vert$, $
\pi _{3}=\pi _{9}^{+}=\left\vert \epsilon _{1}\left( t\right)
\right\rangle \left\langle \epsilon _{2}\left( t\right) \right\vert$, $\pi
_{4}=\pi _{10}^{+}=\left\vert \epsilon _{4}\left( t\right) \right\rangle
\left\langle \epsilon _{3}\left( t\right) \right\vert$, $\pi _{5}=\pi _{11}^{+}=\left\vert \epsilon _{1}\left( t\right)
\right\rangle \left\langle \epsilon _{3}\left( t\right) \right\vert$, $\pi _{6}=\pi _{12}^{+}=\left\vert \epsilon _{1}\left( t\right)\right\rangle \left\langle \epsilon _{4}\left( t\right) \right\vert$, $N_{l}(\epsilon_{ij})=1/\left( e^{\epsilon _{ij}/K_B T_l}-1\right) $.
In the above deduction, the usual Born-Markov approximation and the rotating wave
approximation are performed, and an Ohmic spectral density with infinite
cut-off frequency is also assumed for the heat bath. To evaluate the dynamic
features of the systems under dissipation, we let $\rho_{kl}\left(
t\right) =\left\langle \epsilon _{i}\left( t\right) \right\vert \rho
\left\vert \epsilon _{i}\left( t\right)
\right\rangle $, and $X\left( t\right) =\left[ \rho_{11},\rho
_{22},\rho_{33}\right] ^{T}$, with the relation $\rho_{11}+\rho
_{22}+\rho_{33}+\rho_{44}=1$. Based on these notations, we get the
following dynamic equations that denote the process of thermalization%
\begin{equation}
\dot{X}\left( t\right) =-M\left( t\right) X\left( t\right) +R\left( t\right),
\end{equation}%
and
\begin{equation}
\dot{\rho}_{kl}\left( t\right) =-\left( \eta _{l}+\eta _{k}-i\epsilon
_{lk}\right) \rho_{kl}\left( t\right) ,k\neq l,
\end{equation}%
where $R\left( t\right) =2\left[ \xi_{9},\xi_{11,}-\xi_{12}\right] ^{T}$, $%
\eta _{1}=\xi _{9}+\xi _{11}+\xi_{12},$ $\eta _{2}=\xi _{1}+\xi _{2}+\xi_{3},$ $\eta _{3}=\xi
_{4}+\xi_{5}+\xi _{7},$ $\eta _{4}=\xi _{6}+\xi _{8}+\xi_{10},$ and%
\begin{equation}
M\left( t\right) =2\left[
\begin{array}{ccc}
\xi _{9}+\eta _{2}& \xi _{9}-\xi_{7} & \xi _{9}-\xi _{8}
\\
\xi _{11}-\xi_{1} & \xi _{11}+\eta _{3} & \xi _{11}-\xi _{10}
\\
\xi _{12}-\xi_{2}&\xi_{12}-\xi _{4} &
\xi_{12}+\eta _{3}%
\end{array}%
\right] .
\end{equation}

Based on these equations, we can obtain the elements of density matrix in the original space
\begin{eqnarray}
\sigma_{ij}\equiv Tr[\rho|\eta_j\rangle\langle\eta_i|]=\sum_{k=1}^4\sum_{l=1}^4u_{jk}u_{il}^*\rho_{lk}.
\end{eqnarray}

In the following section, we will show some numerical results of the dynamical properties of the model based on Eq. (15).
\subsection{Excited by a single Gaussian type pulse}

\begin{figure}[tbph]
\centering \includegraphics[width=8.0cm]{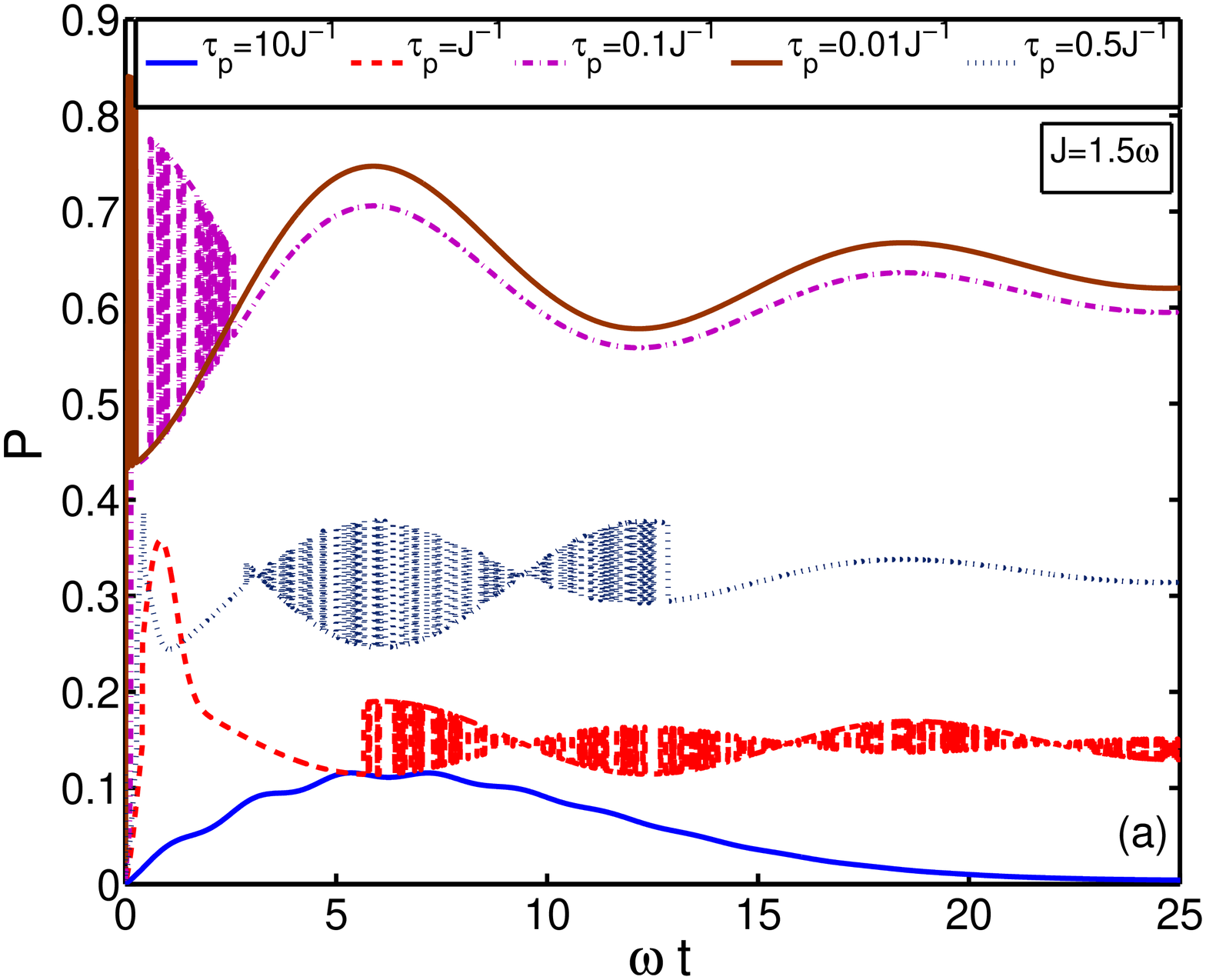}%
\includegraphics[width=8.0cm]{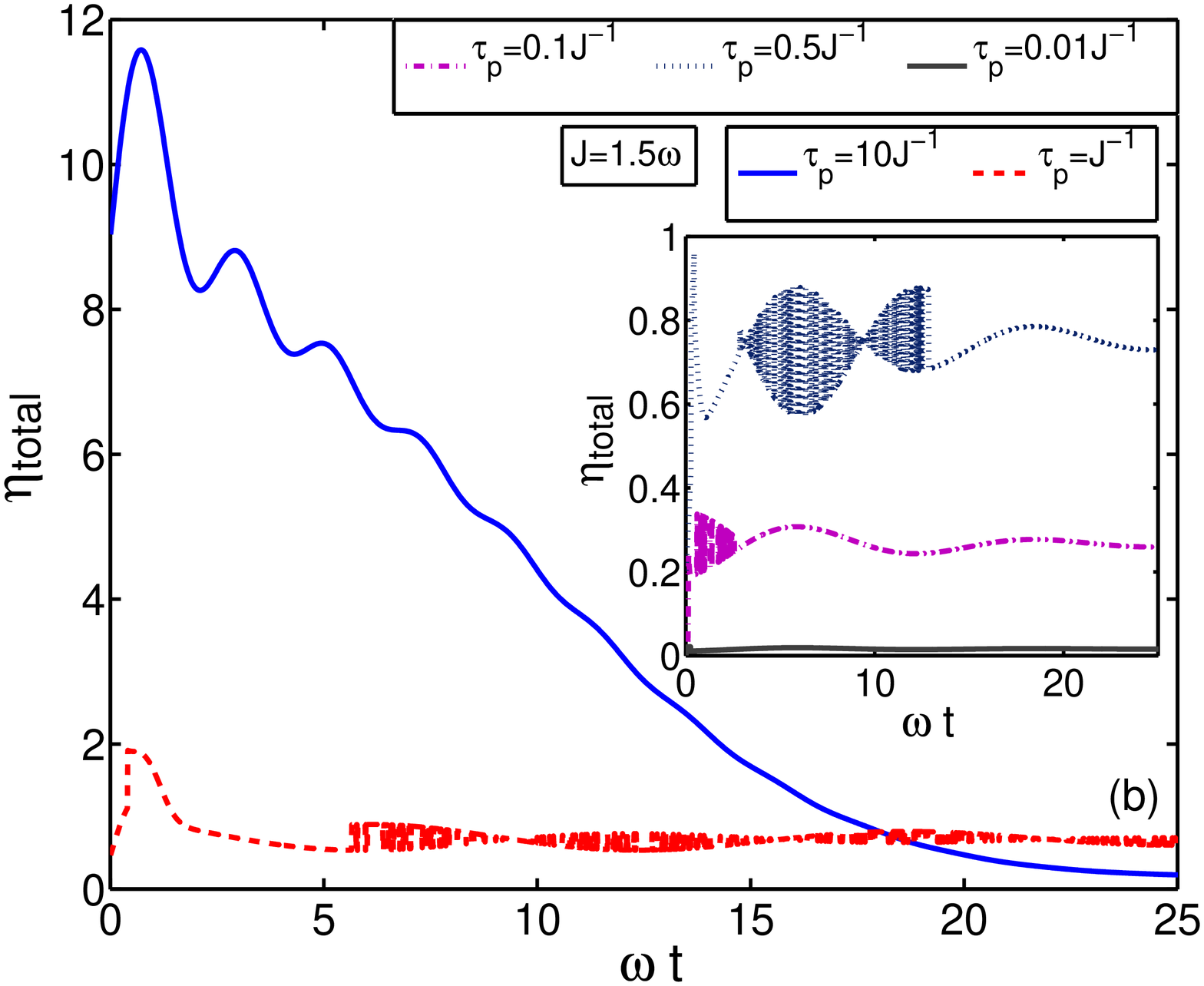}
\caption{(color online) Probability $P$ that the acceptor is in the excited state and the parameter of total efficiency $\eta_{total}$ versus time $\omega t$ are plotted. We set $\omega_1=\omega_2=\omega$, $J=1.5\omega$, $\kappa_1=\kappa_2=0.1\omega$, $T_1=T_2=0.1\omega$. The same parameters are also chosen in the following figures unless special mention.}
\end{figure}

We first suppose that the input pulse is a Gaussian type pulse, which has the same form as we mentioned previously. In order to compare the input and output energy in our numerical results, we assume that the amplitude of input pulse is $E_0=\omega_1$.

In Fig. 1a and Fig. 1b, we plot the probability $P$ that the acceptor pigment is in the excited state and the associated $\eta_{total}$ versus time, respectively. We let $\omega_1=\omega_2=\omega$ \cite{[41]}, and set $\omega=1$ as the calculation unit. The largest width of the pulse is chosen as $\tau_p=10J^{-1}$ (Solid and blue line), in which $J^{-1}$ is used to valuate time scale of state exchanging between the two pigments. We set $\tau_p=0.01J^{-1}$ (solid and brown line) to simulate a $\delta$ pulse. We find from the figures that both of $P$ and $\eta_{total}$ oscillate with time and approach to saturation. For the case that $\tau_p$ are small, oscillation represents that the excited states and excitation energy transfer between the two pigments. But for the case of $\tau_p=10J^{-1}$, we easily find that the dynamics of $P$ is in connection with the processes of excitation. In the case of inputing a $\delta$ pulse, the acceptor is always in the excited states but the dimer has the smallest total efficiency $\eta_{total}$, which means that higher probability of the acceptor being in the excited state is not equivalent to successful energy absorbing and transferring in a photochemical reaction process. Figure 2 shows that the saturation values of $P$ are decreasing with $\tau_p$ increasing, but there has a optimum intervals of $\tau_p$ for the saturation values of $\eta_{total}$.

\begin{figure}[tbph]
\centering \includegraphics[width=10.0cm]{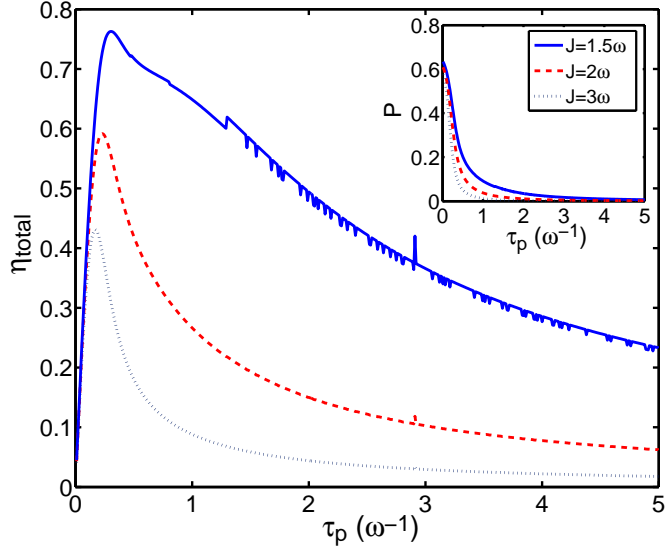}
\caption{(color online) The saturation values of $\eta_{total}$ and $P$ (Inset) with different $J$ versus $\tau_p$ are plotted.}
\end{figure}

We also draw the dynamics of quantum concurrence between the two pigments in Fig. 3. We observe from the figures that quantum entanglement can be produced in the dimer when the donor is excited by a pulse, and find that shorter temporal pulse will induce larger entanglement.
\begin{figure}[tbph]
\centering \includegraphics[width=10.0cm]{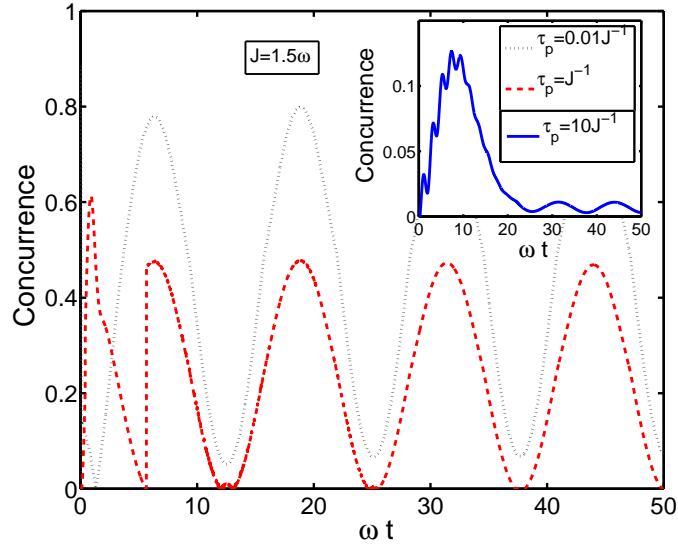}
\caption{(color online) Quantum concurrence between the two pigments versus time $\omega t$ are plotted.}
\end{figure}

\begin{figure}[tbph]
\centering \includegraphics[width=8.0cm]{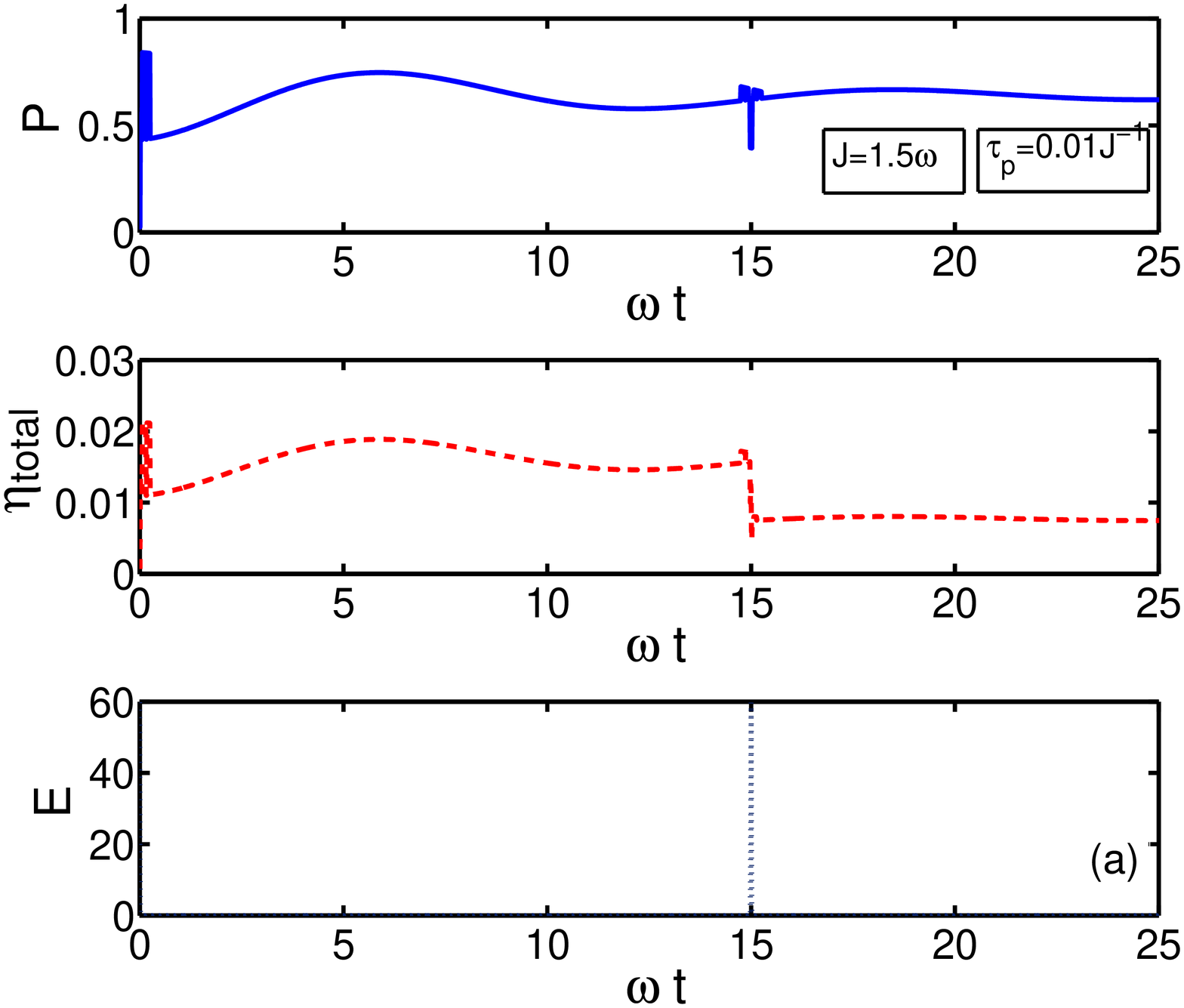}%
\includegraphics[width=8.0cm]{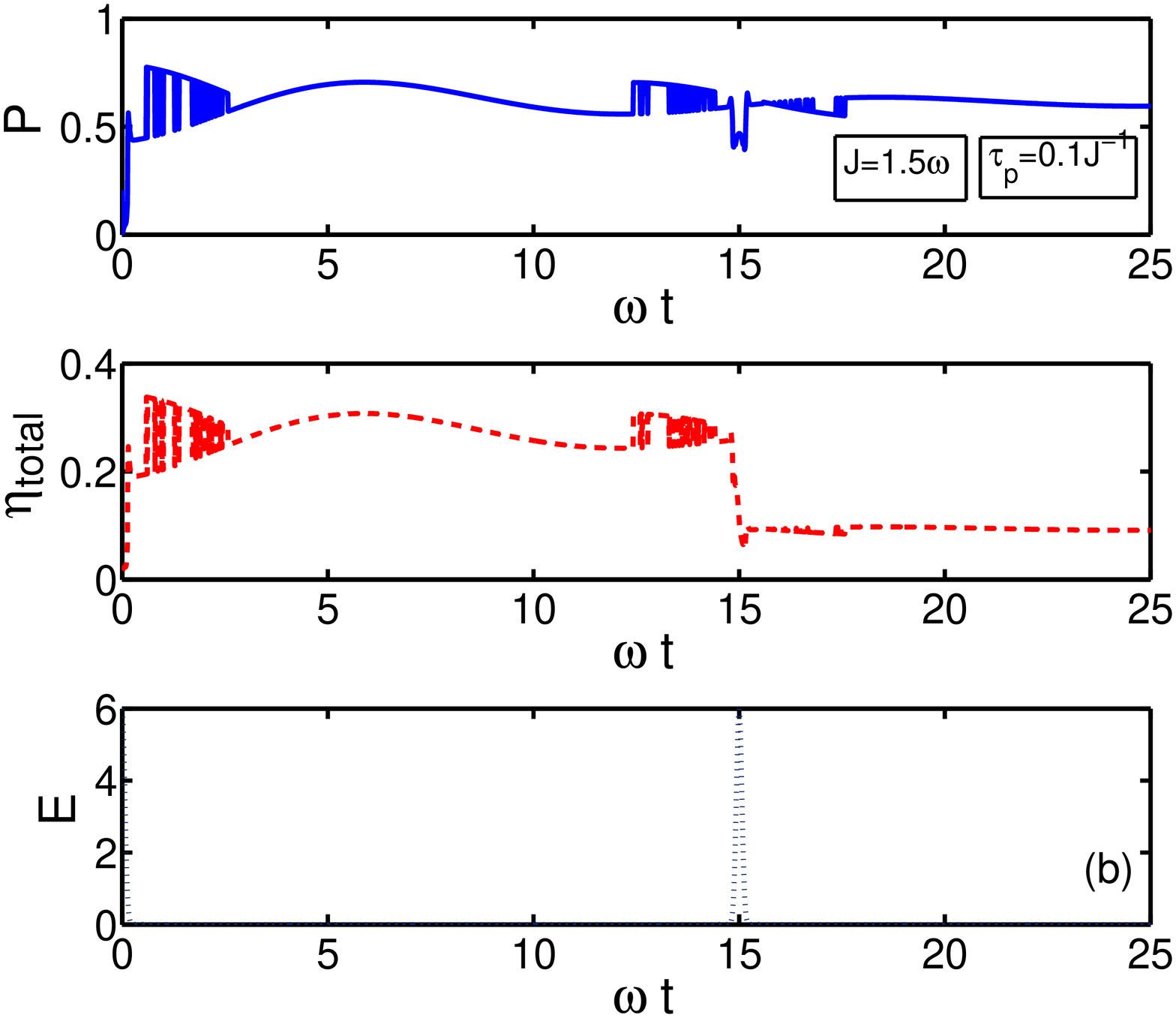}
\caption{(color online) Probability $P$ that the acceptor is in the excited state and the parameter of total efficiency $\eta_{total}$ versus time $\omega t$ are plotted. We set width of the pulse are $\tau_p=0.01J^{-1}$ and $\tau_p=0.1J^{-1}$ in the Fig. 4a and Fig. 4b, respectively.}
\end{figure}

\subsection{Excited by sequential Gaussian type pulses}
In the above, we have studied the dynamics of a dimer being excited by a single Gaussian pulse. In order to understand how temporal properties of an input pulse affects the quantum dynamics of the dimer, we now consider some more complex cases. 

In the figures from Fig. 4a to Fig. 6a, we consider the case that the donor pigment being excited sequentially by two same Gaussian type pulses, the central time of the first pulse is at $\omega t=0$, and the central time of the second pulse is in $\omega t=15$. In Fig. 4a and Fig. 4b, we set the width of the pulse are $\tau_p=0.01J^{-1}$ and $\tau_p=0.1J^{-1}$, respectively. We find that in both cases, comparing with the associated Fig. 1a and Fig. 1b, probability $P$ has only a slit sharp down during the second pulse acting, and $\eta_{total}$ has a sudden decrease.  
However, when we set $\tau_p=0.5J^{-1}$ and $\tau_p=J^{-1}$ in Fig. 5a and Fig. 5b, respectively, we find contrary phenomena, $i. e.$ probability $P$ has a upward change during the second pulse acting on, and similar changes to $\eta_{total}$ with decreasing finally. There should be two physical mechanisms to understand the phenomena in Fig. 4 and Fig. 5. In the first aspect, we observe that probability $P$ is always larger than $0.5$ when the second pulse is absent in the cases of $\tau_p=0.01J^{-1}$ and $\tau_p=0.1J^{-1}$, which means that the acceptor pigment is alway in its excited state after the first pulse acting on, so the second pulse will induce stimulated radiation with larger probability than that of stimulated absorption. Second, one need also to note the fact that excited state absorption can also deplete population in the single exciton manifold (i.e., population in the acceptor). This means that the coherent effects from the sequential multi-pulses in the acceptor pigment take place during a short timescale, and thus deplete population. It will induce contrary results in the cases of $\tau_p=0.5J^{-1}$ and $\tau_p=J^{-1}$ because that probability $P$ is always smaller than $0.5$. 
\begin{figure}[tbph]
\centering\includegraphics[width=8.0cm]{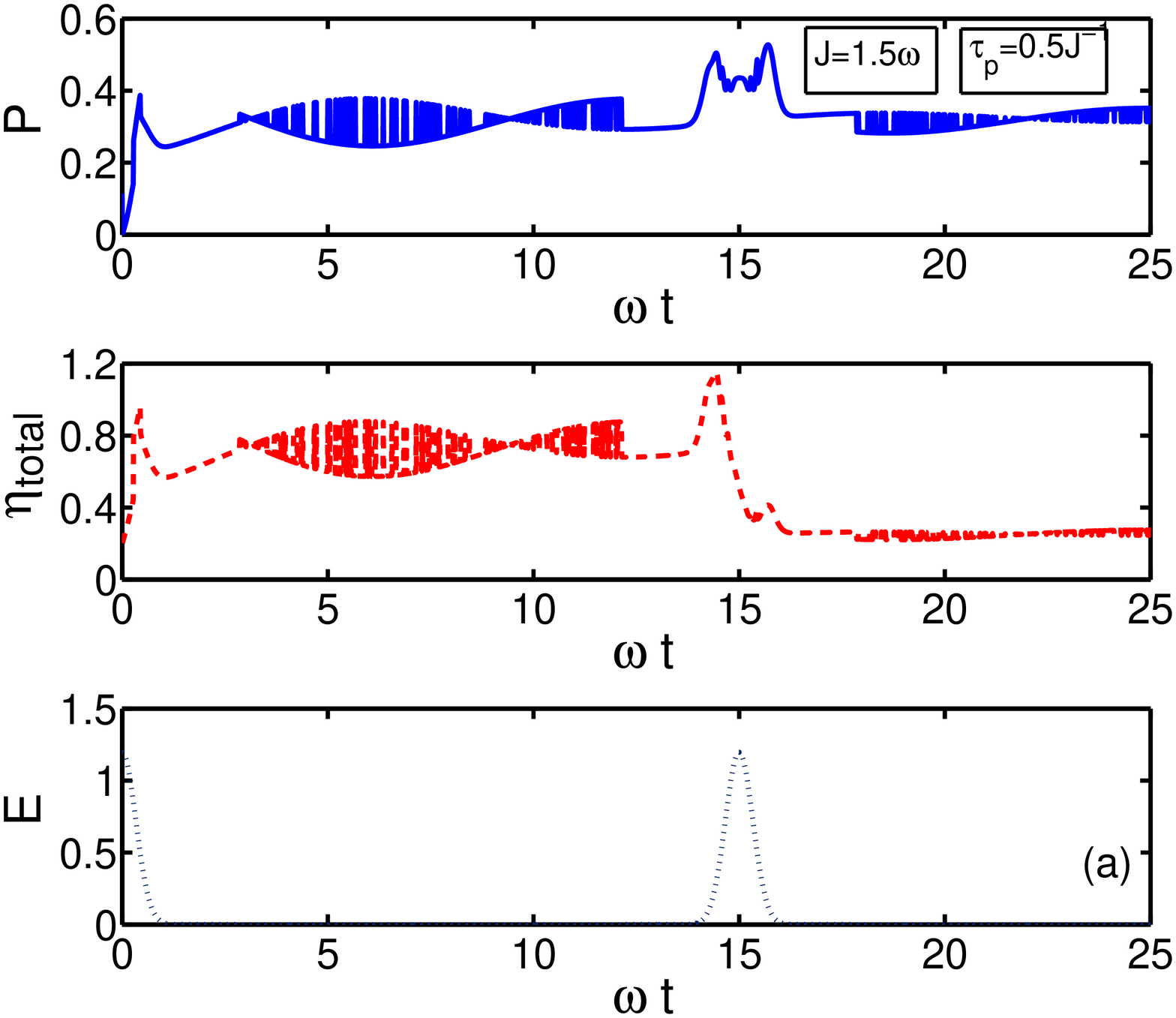}%
\includegraphics[width=8.0cm]{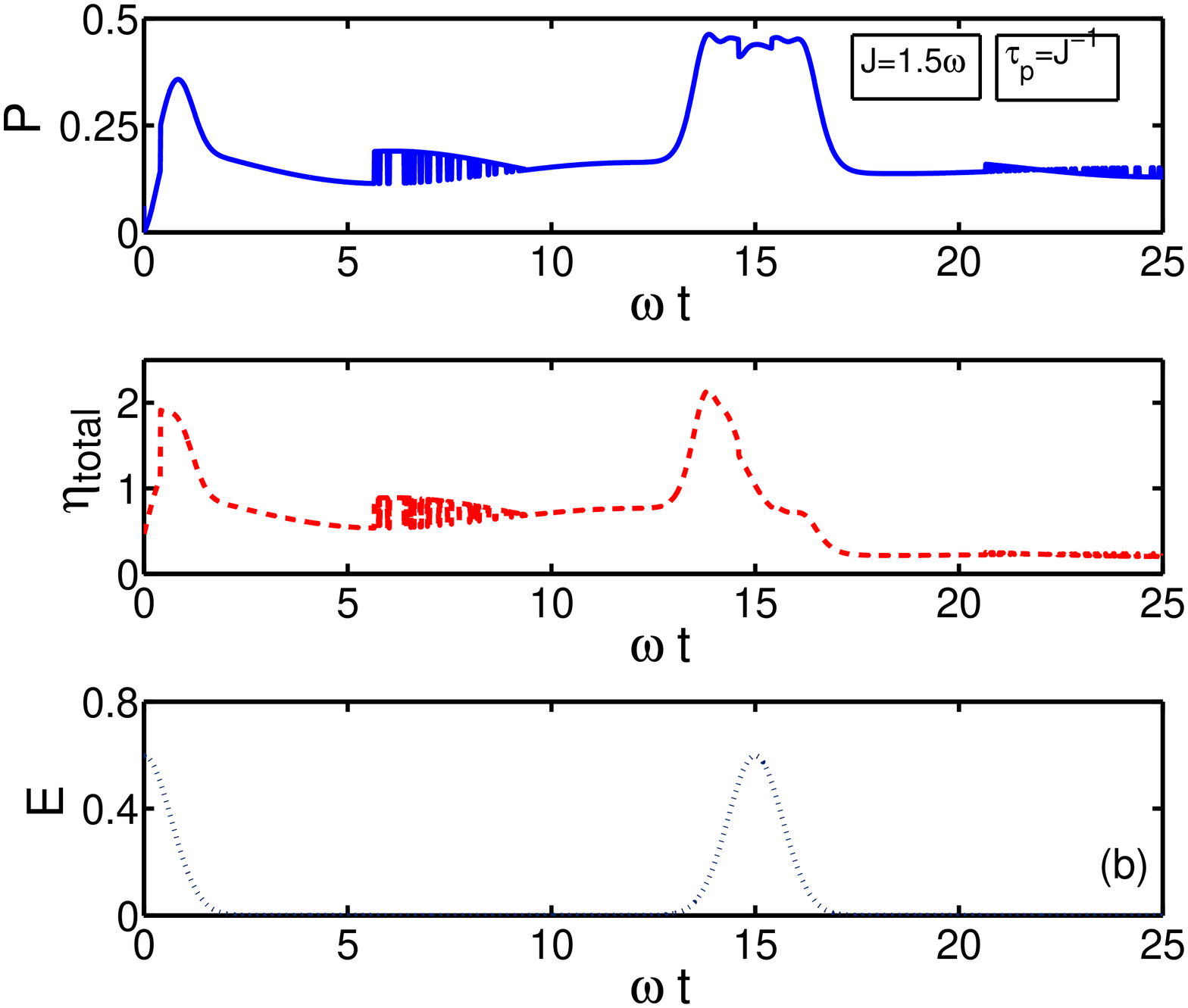}
\caption{(color online) Probability $P$ that the acceptor is in the excited state and the parameter of total efficiency $\eta_{total}$ versus time $\omega t$ are plotted. We set the width of pulses as $\tau_p=0.5J^{-1}$ and $\tau_p=J^{-1}$ in the Fig. 5a and Fig. 5b, respectively.}
\end{figure}
\begin{figure}[tbph]
\centering\includegraphics[width=8.0cm]{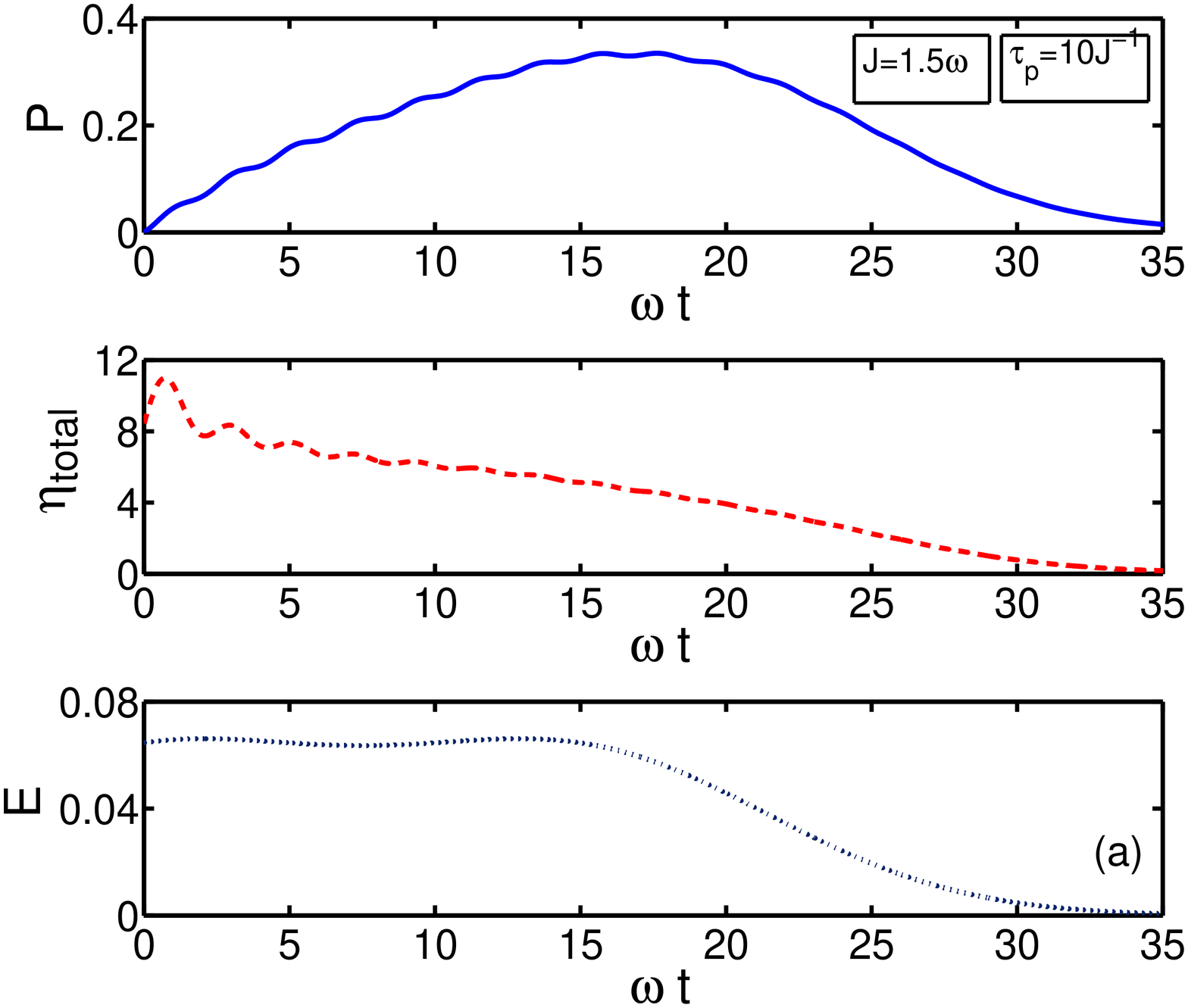}%
\includegraphics[width=8.0cm]{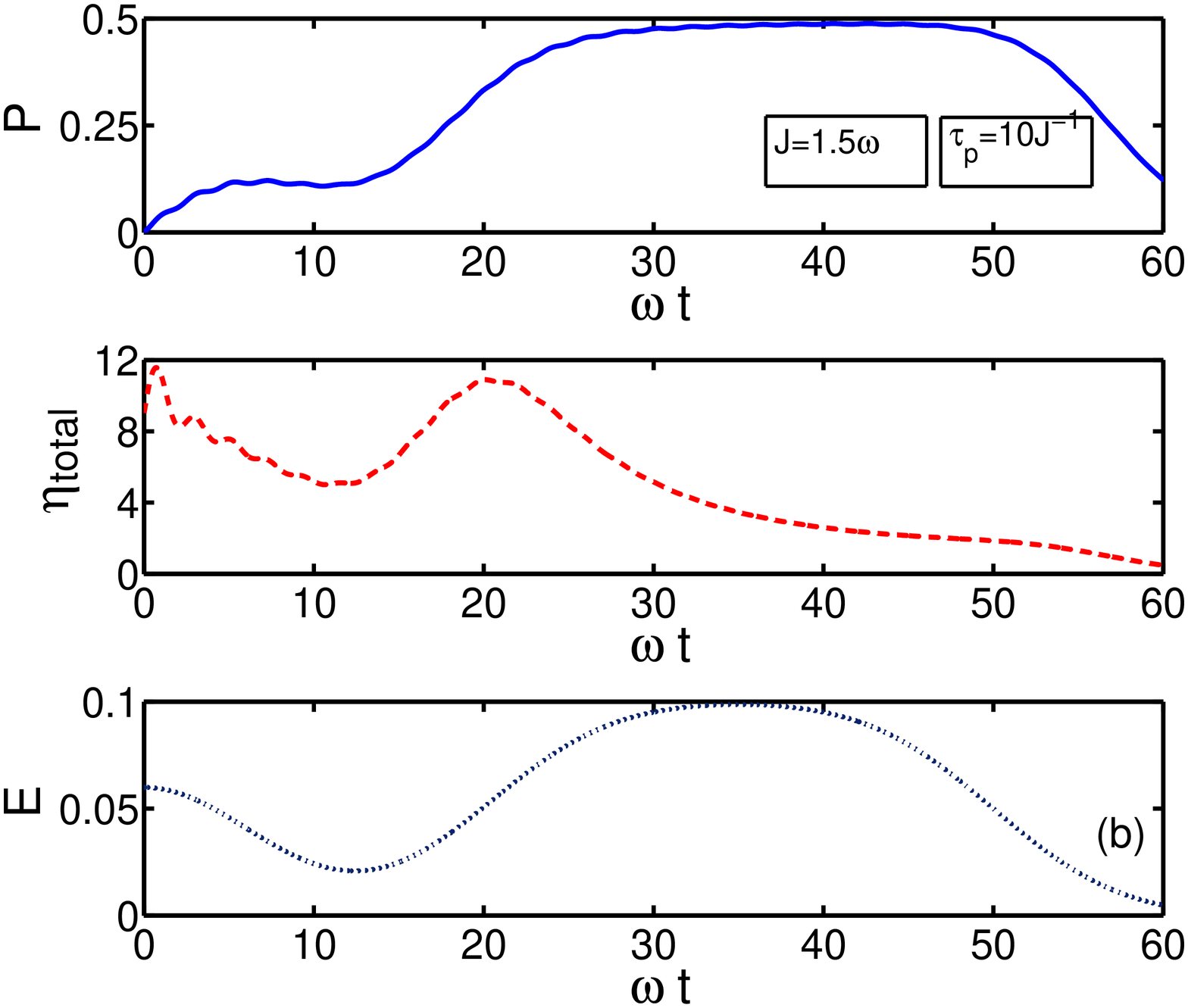}
\caption{(color online) Probability $P$ that the acceptor is in the excited state and the parameter of total efficiency $\eta_{total}$ versus time $\omega t$ are plotted. In the Fig. 6b, four sequential pulses are acted on the donor pigment, the width of pulses are all $\tau_p=10J^{-1}$, and the central action times are $\omega t=0,~25,~35,~45$, respectively.}
\end{figure}

In Fig. 6a, we set $\tau_p=10J^{-1}$ and produce a near constant pulse during time from $\omega t=0$ to $\omega t=15$, that makes a continuous enhancement of probability $P$ comparing with Fig. 1a but the total efficiency $\eta_{total}$ decreases at all the time. In Fig. 6b, we show the case that four sequential pulses are acted on the dimer, width of pulses are all $\tau_p=10J^{-1}$, and the central action times are $\omega t=0,~25,~35,~45$, respectively. We increase time interval of the first and the second pulse to $25\omega t$, and produce a near constant $P$ during $\omega t=25$ to $\omega t=50$, by which we realize controlling populations of the acceptor pigment simply by adjusting the temporal shapes of the input pulses. We also obtain a extreme value of $\eta_{total}$ near $\omega t=20$. But as we have mentioned previously, the values of $P$ are always not larger than $0.5$. 

\begin{figure}[tbph]
\centering \includegraphics[width=8.0cm]{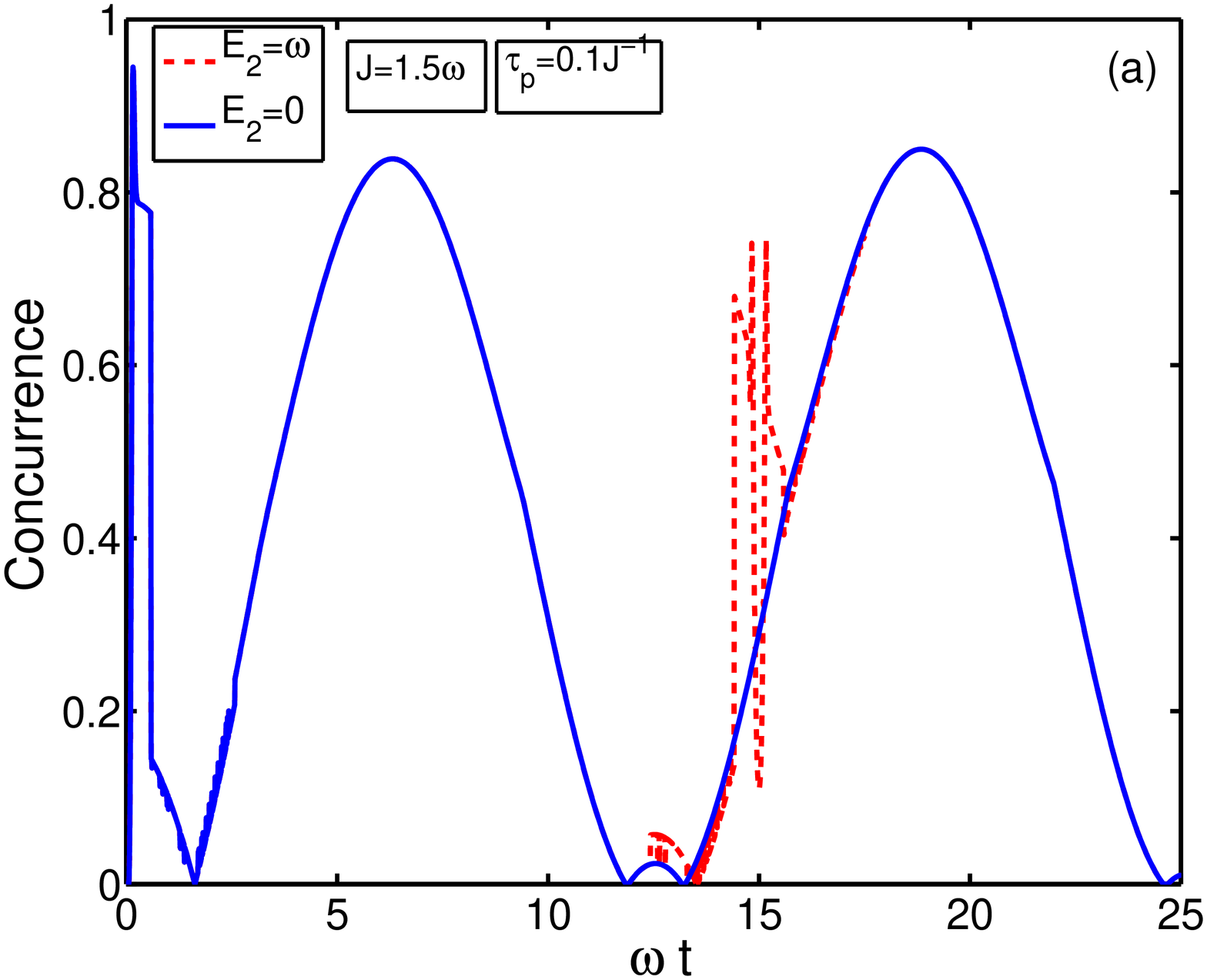}%
\includegraphics[width=8.0cm]{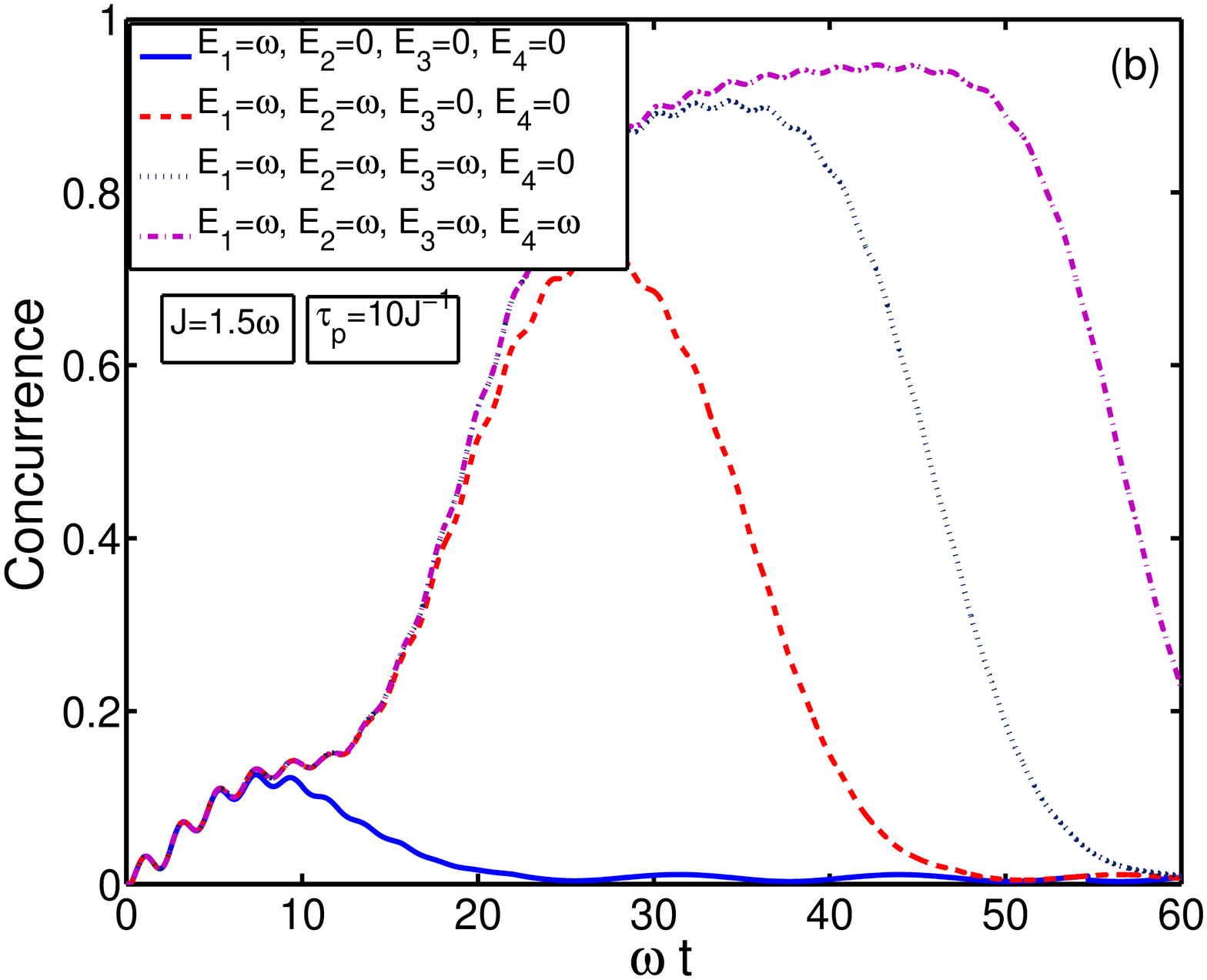}
\caption{(color online) Quantum concurrence between the two pigments versus time $\omega t$ are plotted. In the Fig. 7b, four pulses are turned on sequentially, the width of pulses are all $\tau_p=10J^{-1}$, and the central action times are $\omega t=0,~25,~35,~45$, respectively.}
\end{figure}

Finally, we plot quantum concurrence between the two pigments when the donor pigment is excited by multi-pulses sequentially. In Fig. 7a, we consider the case that two sequential pulses act on, where the central time of the first pulse is at $\omega t=0$, the central time of the second pulse is at $\omega t=15$, and $\tau_p=0.1J^{-1}$. We find there is only a disturb of quantum concurrence when the second pulse acts on. However, in Fig. 7b, quantum concurrence is largely enhanced when the second pulse, the third pulse, and the fourth pulse turn on sequentially, where we set $\tau_p=10J^{-1}$ and the central times of the four pulses are same as shown in Fig. 6b. Comparing with the results shown in Fig. 4 and Fig.5, where the probability $P$ can only be enhanced to a maximum value $0.5$, we find quantum concurrence can be enhanced to very large values. It is because that quantum coherence between the two pigments can be continuously induced no matter the system is in the processes of stimulated emission or stimulated absorption.

\section{Conclusions}

In summary, we have studied controlling excitation and coherent transfer in a dimer. In our model, energy transferring from the donor to the acceptor takes place during the processes that the donor is being excited. The model can be applied to find physical mechanism of the basic processes of energy absorption and transferring for photosynthesis. We mainly investigated how temporal shapes of the input pulses affect the population behavior of the acceptor and quantum concurrence of the dimer. We find that high probability of the acceptor being excited can be obtained with very sharp pulse but with a very low total efficiency of energy absorption and transferring. The total efficiency depends on the temporal shape of the input pulses. When the dimer is excited by sequential multi-pulses, there are two physical mechanisms to determine probability of the acceptor being excited; one is the coherent effects from the sequential multi-pulses in the acceptor pigment during a short timescale; another is the population conditions of the acceptor before being pumped again, $i.e.,$ the later pulses will induce the acceptor pigment being stimulated emission or stimulated absorption. Our results also show that high degree quantum concurrence of the dimer can be obtained by controlling temporal shape of the input pulses. Our studies may contribute to the fundamental research of artificial photosynthetic units.

\section*{Acknowledgments}
This work is supported by the National Natural Science Foundation of China (grant No. 11174233, 11004158), the Special Prophase Project on the National Basic Research Program of China (grant 2011CB311807), and the Fundamental Research Funds for the Central Universities.


\begin{thebibliography}{99}
\bibitem{[41]} We also studied the case of $\omega_1\neq \omega_2$, and found the same main results with the present case.
\end{thebibliography}
\end{document}